# A hybrid six-dimensional muon cooling channel using gas filled rf cavities

**Diktys Stratakis,**[a]

[a] *Fermi National Accelerator Laboratory,*
 *Batavia,IL 60510 USA*
 *E-mail*: diktys@fnal.gov

ABSTRACT: An alternative cooling approach to prevent rf breakdown in magnetic fields is described that simultaneously reduces all six phase-space dimensions of a muon beam. In this process, cooling is accomplished by reducing the beam momentum through ionization energy loss in discrete absorbers and replenishing the momentum loss only in the longitudinal direction through gas-filled rf cavities. The advantage of gas filled cavities is that they can run at high gradients in magnetic fields without breakdown. With this approach, we show that our channel can achieve a decrease of the 6-dimensional phase-space volume by several orders of magnitude. With the aid of numerical simulations, we demonstrate that the transmission of our proposed channel is comparable to that of an equivalent channel with vacuum rf cavities. Finally, we discuss the sensitivity of the channel performance to the choice of gas and operating pressure.

# Contents

## 1. Introduction

Cooling techniques are used to reduce the phase-space of beams in particle accelerators. These techniques reduce the relative motion of particles in a bunch by acting on individual or finite sized groups of particles. Electron [1] and stochastic cooling [2] are both well-established techniques and have been used to cool antiprotons [3, 4]. However, muons have a short lifetime and must be cooled quickly for a muon collider [5] or a neutrino factory [6].

Ionization cooling [7] appears to be the only promising mechanism that can provide a sufficiently short damping time. Ionization cooling is achieved by reducing the beam momentum through ionization energy loss in absorbers, which interact less with muons than with antiprotons, and replenishing the momentum only in the longitudinal direction using rf cavities. This mechanism can effectively reduce the transverse phase space of a beam in an analogous way as radiation damping does to an electron beam [8]. However, this mechanism does not cool the longitudinal momentum spread, because at 200 MeV/c where transverse cooling works well, the energy loss is falling with increasing momentum. To obtain longitudinal cooling, dispersion is introduced in order to spatially separate muons of different longitudinal momenta, and wedge-shaped absorbers used to reduce the momentum spread. Such a longitudinal cooling scheme is called "emittance exchange" because the longitudinal cooling is achieved at the expense of transverse heating or a reduced transverse cooling rate.

While ionization cooling seems a straightforward method from a theoretical point of view, in reality it requires extensive simulation and hardware development for its optimization. Over the past decade much progress has been made in the design and simulation of lattices that use emittance exchange in order to cool all six dimensions of a beam's phase space [9-13]. In a recent study [14], it was shown numerically and theoretically that by using a 12-stage rectilinear channel the 6D emittance can be cooled by more than 5 orders of magnitude. A key requirement of that design is the use of vacuum rf cavities in 5 T or greater magnetic fields. However, operational experience (specifically in the Fermilab Mucool Test Area) as well as numerical studies [15] suggest that the performance of a normal conducting cavity may degrade when the cavity is operated in such strong magnetic fields. The magnetic fields cause rf cavity breakdown at high gradients. On the other hand, recent experiments using a cavity filled with high-pressure (hp) hydrogen gas show no degradation in maximum gradient in this configuration [16, 17]. For this reason, it seems prudent to investigate options for using hp rf in a 6D ionization cooling channel.

The goal of this work is to present an alternative cooling scheme that simultaneously reduces the size of a charged particle beam in all six phase-space dimensions. Our lattice design is based on a rectilinear channel [14] with the key difference that we examine a hybrid approach [18] wherein we incorporate gas-filled pressurized cavities along with discrete lithium hydride (LiH) absorbers. With the aid of simulation codes, we compare its performance against the performance of an equivalent channel with vacuum rf cavities. We show that a hybrid channel not only maintains a transmission comparable to that of a vacuum channel, but also achieves a comparable 6D emittance cooling rate. Finally, we explore the sensitivity of the performance to gas pressure.

The outline of this paper is as follows: In Sec. II, we give an overview of proposed alternative 6D cooling schemes that rely on vacuum rf cavities. In Sec. III, we provide details of the design parameters of our proposed hybrid channel. Next, in Sec. IV we report the results of simulations modeling the aforementioned channel and examine its sensitivity to various gas parameters. Finally, we present conclusions in Sec. V.

## 2. Rectilinear channel

In the scheme considered in Ref. 11 (often called the Guggenheim; see Fig. 1), the lattice is bent into a helix with curvature corresponding to that generated by the dipole field component. The channel consists of a series of identical cells with two or four solenoidal magnets in each cell with opposite polarities to provide transverse focusing. The coils (yellow) are not evenly spaced; those on either side of the wedge absorber (magenta) are closer together in order to increase the focusing at the absorber and thus minimize the beta function at that location. When the beta function is small, the effect of multiple scattering in the absorber becomes a smaller fraction of the overall angular beam divergence and thus a smaller contribution to the transverse emittance. The amount of transverse cooling vs. emittance exchange can be adjusted by changing the opening angle and transverse location of the wedge. A series of rf cavities (dark red) are used to restore the momentum along the longitudinal axis. The dispersion necessary for emittance exchange is provided from the bend field generated by tilting the axes of the solenoids upward or downward from the orbital midplane.

Simulations have shown that a suitable sequence of such helices, with multiple stages using different cell lengths, focusing fields, and rf frequencies, can provide a three orders of magnitude reduction of the normalized phase-space volume with a transmission above 50%. Although, this case would appear to be practical for the early stages of 6D cooling, it would become increasingly difficult as the helix radii decrease in the later stages. An added complication is that stray fields from one turn can influence those before and after, causing the beam to become distorted.

In an alternative scheme proposed by Balbekov [19] (see Fig 2), essentially the same cells from a Guggenheim, including their coil tilts and resulting upward dipole fields, are laid out in straight (rectilinear) geometry. The solenoid focusing is so strong compared with the dipole deflections that the closed orbits are merely displaced laterally, but continue down the now straight lattice. Numerical simulations showed [14] that its cooling performance was essentially the same as that of a Guggenheim channel. Therefore, the rectilinear scheme will be considered our baseline lattice and will be analyzed in more detail in the next section.

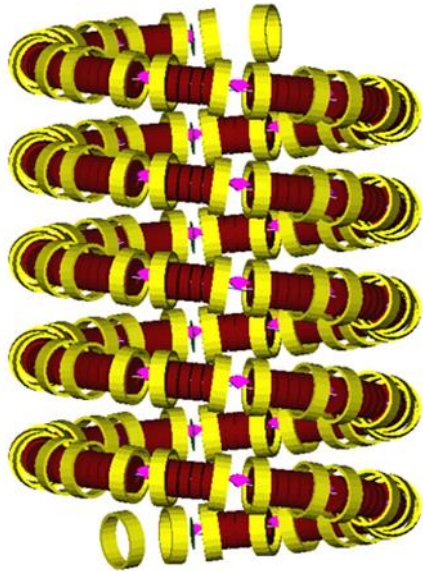

**Figure 1:** Helical cooling channel commonly known as Guggenheim [11]; 5 turn slice shown.

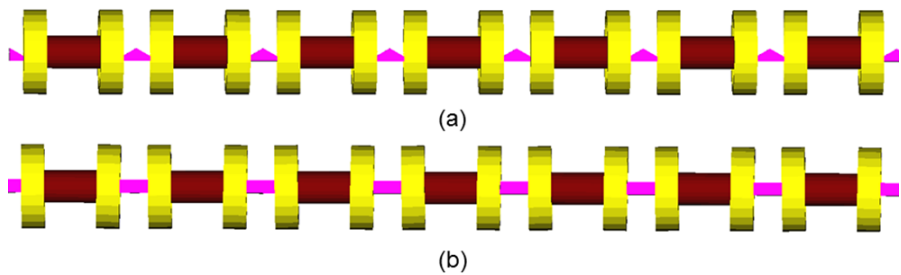

**Figure 2:** Conceptual design of a rectilinear channel [19]: (a) Top view; (b) Side view.

## 3. Hybrid solution

A key requirement of a vacuum ionization cooling channel is that rf cavities operate in 5 T or greater magnetic fields. An rf cavity filled with hp hydrogen gas has been proposed as a potential solution to protect the cavities from breaking down in such high fields. More precisely, experiments [16] have demonstrated that a breakdown gradient of 65.5 MV/m could be achieved in a 3 T magnetic field with 70 atm hydrogen gas. This 805 MHz cavity was later successfully tested with an ionizing beam in the parameter range of interest [17].

In order to incorporate the aforementioned gas filled cavities into a rectilinear ionization cooling channel we investigate a hybrid approach. Our proposed scheme will use hp gas to protect the cavity, along with discrete LiH absorbers to provide the majority of the energy loss. Fig. 3 shows a schematic illustration of a cooling cell with vacuum cavities [Fig. 3(a)] and gas filled cavities [Fig. 3(b)]. Since our primary purpose is to protect the cavity from the effects of high magnetic field we use only modest gas pressure to accomplish this task. Based on the measurement in Ref. 16 at 805 MHz, we expect that a pressure of 34 atm at room temperature will be satisfactory. Eventually this parameter needs to confirmed in both 325 MHz and 650 MHz rf cavities which are commonly used in ionization cooling channel designs. At our specified pressure, to compensate for the energy loss in the gas we slightly reduce the length of the LiH absorber by 5%.

Here it is important to emphasize that there are key differences between our proposed scheme and gas helical cooling channels with hp rf cavities as studied in Ref. [13]. In a gas helical cooling channel, the energy loss is distributed uniformly throughout the channel using 160 atm hydrogen, rather than localized at discrete absorber regions. The addition of 160 atm hydrogen to the rf cavity likely increases the maximum allowable gradient somewhat. On the other hand, maintaining the discrete LiH absorbers to provide the majority of the energy loss requires only a modest pressure which in our case is near 30-40 atm. Cooling is maximized for two reasons. First, it allows stronger magnetic focusing in the limited absorber regions. Second, the effect of multiple scattering in the absorber is now a smaller fraction of the overall angular beam divergence and thus a smaller contribution to the transverse emittance.

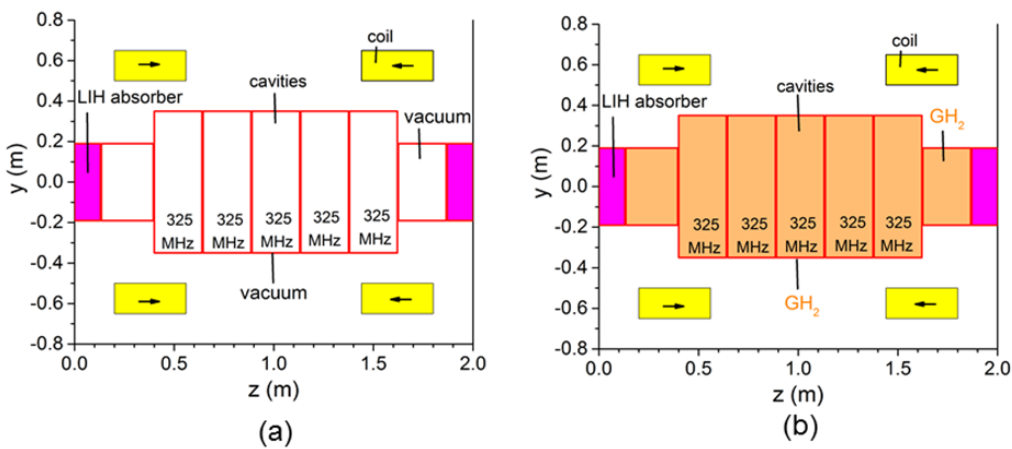

**Figure 3:** Schematic illustration of a cooling channel cell for stage 2: (a) channel with vacuum rf cavities; (b) proposed channel with gas filled cavities.

Recent studies [12] showed numerically that good cooling efficiency requires the channel to be tapered. In this scheme, parameters such as cell length, focusing strength, rf frequency and gradient progressively change from stage to stage based on the emittance reduction rate and transmission to maintain a beam emittance that is always larger than the equilibrium emittance. As a result, each subsequent stage will have a minimum beta [20] lower than the previous one. In the first stage of the tapered channel the focusing will be relatively weak to avoid excessive angular divergence that can arise from the large transverse emittance of the initial muon beam. However, the weak focusing implies that the beta function and thus the equilibrium emittance are relatively large and the transverse cooling weakens as the equilibrium emittance is approached. To avoid this, this stage is terminated and we couple into the next stage, which has a lower beta. This is achieved by simultaneously scaling down the cell dimensions and raising the strength of the on-axis solenoidal field. The result is a piecewise constant multi-stage channel of cells similar to the one shown in Fig. 3. As we will demonstrate in Sec. IV, 9 tapered stages are enough to cool towards the baseline requirements of a Muon Collider. The required lattice parameters are summarized in more detail in Table 1.

Fig. 4(a) exhibits the beta function (at central momentum) and the axial peak field at different stages. The beta function varies from 39.7 cm to 3.3 cm, while the on-axis magnetic field increases from 2.8 T to 13.4 T. The minimum normalized transverse emittance that can be achieved for a given absorber in a given focusing field is:

$$\varepsilon_{N,min} \cong \frac{\beta_T (E_s)^2}{2g_t \beta m_\mu c^2 L_R (dE_\mu/ds)} \quad (3.1)$$

where $E_\mu$ is the muon energy, $\beta_T$ is the transverse betatron function within a discrete absorber, $g_t$ is the transverse partition number, $L_R$ is the radiation length of the material, and $E_s$ is the characteristic scattering energy (~13.6 MeV). Fig. 4(b) shows the transverse equilibrium emittance at different stages as calculated from Eq. (3.1). One can see that, at least in theory, our proposed scheme can cool the beam to the required Muon Collider [21] transverse emittance of ~ 0.30 mm. A technical challenge may arise as the operating current in the conductor should not exceed the critical current corresponding to the peak field at the coil. Note that there is an increase of the magnet operating current with stage number. This is a direct consequence of the low beta lattice design which is needed to cool towards micron scale emittances. Most challenging is the last stage where the solenoids are expected to deliver a field on-axis near 14 T. A detailed feasibility study for this last stage can be found in Ref. [22]. The analysis highlighted that for stable operation $Nb_3Sn$ coils at a 1.9 K operating temperature are preferred.

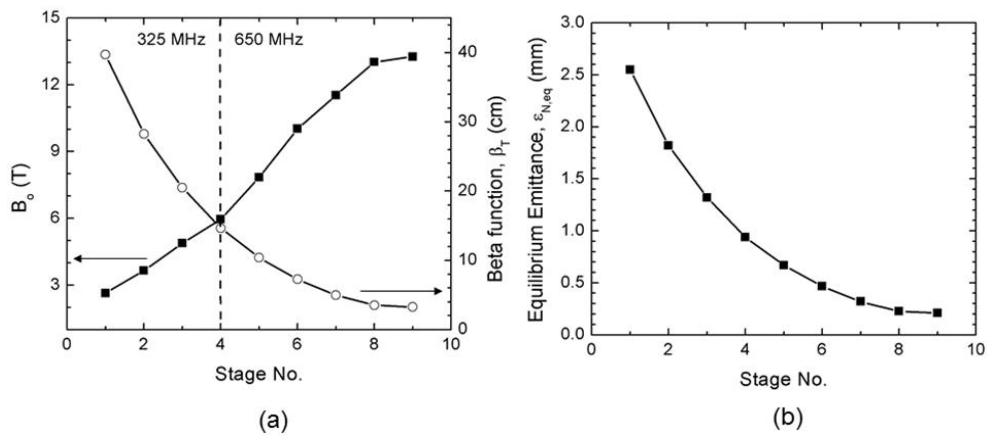

**Figure 4:** (a) Peak magnetic field on beam axis and corresponding transverse beta function vs. stage number, and (b) equilibrium emittance vs. stage number.

## 4. Simulation results

The performance of the cooling channel was simulated using the ICOOL code [23]. The code includes all relevant physical processes, e.g. energy loss, straggling, multiple scattering, and muon decay. For each stage, we generated 3D cylindrical field maps by superimposing the fields from all solenoids in the cell and its neighbor cells. The rf cavities were modeled using cylindrical pillboxes running in the TM010 mode and a reference particle was used to determine each cavity's relative phase. The absorber material was lithium hydride for all stages. The normalized emittance has been obtained using ECALC9f [24], an emittance calculation program customarily employed by the Muon Accelerator Program [25]. In the calculation, a factor $1/m_\mu c$ was introduced so as to express the longitudinal emittance in units of length.

The input beam in the simulations has a normalized transverse emittance of 5.10 mm and a normalized longitudinal emittance of 10.02 mm, while the average longitudinal momentum is 208 MeV/c. Those parameters closely resemble the baseline distribution of a muon beam before the final 6D cooling sequence of a Muon Collider [12]. We tracked 56,000 particles and included muon decay. Note that average momentum does not remain constant along the channel but is reduced by ~ 10-15MeV/c in the final stages in order to reduce the value of the transverse beta function. This is because in a solenoidal channel, the beta function is proportional to the momentum and inversely proportional to the axial field. Therefore, for a given strength magnet we can focus the beam more strongly if we lower the momentum.

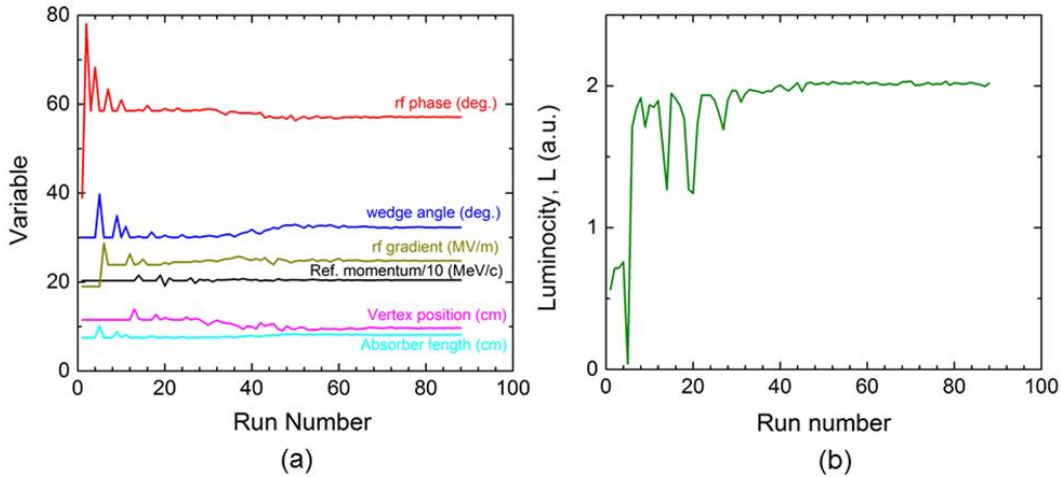

**Figure 5:** Multi-dimensional scans wherein we varied key lattice parameters such as the rf phase, rf gradient, absorber thickness, absorber angle, and rf gradient in order to find the peak luminosity.

For all cooling scenarios examined, we performed multi-dimensional scans wherein we varied key parameters such as the rf phase, rf gradient, absorber thickness, absorber vertex position and absorber angle in order to find the peak muon yield for each variable. Using the "Nelder-Mead" algorithm [26] our goal was to maximize the beam luminosity [27], $L \sim T^2/(\varepsilon_T \varepsilon_L)$ where $T$ is the transmission of the lattice and $\varepsilon_T$, $\varepsilon_L$ are the transverse and longitudinal beam emittances, respectively. In Fig. 5 we show an example of the sensitivity of the performance as we vary the aforementioned parameters for one stage of the channel.

Fig. 6 displays the transmission, transverse emittance and longitudinal emittance vs. distance for a channel with vacuum cavities and our proposed hybrid channel. The dashed curve shows the transmission of muons with the decay option enabled in the simulation. In the case of the hybrid channel [Fig. 6(b)] after a distance of 515 m (9 Stages) the 6D emittance has fallen by a factor of 1000 with a transmission of 35% (52% with decays disabled). In addition, at the end of the channel the transverse emittance has decreased by a factor of ~17, while the longitudinal emittance has decreased by more than a factor of ~6. Note that a transverse emittance $\varepsilon_T \approx 0.3$ mm is the baseline requirement for a Muon Collider at the end of the 6D cooling sequence [21]. We can conclude from the results in Fig. 6(b), that 9 Stages are enough to fulfill this requirement since the transverse emittance of the final beam at z = 515 m is 0.3 mm. Clearly, the channel achieves virtually identical emittances and transmission as the vacuum rf channel [Fig. 6(a)], the only difference being 10 m of added length. Note that the influence of isolation windows is not considered in the analysis. Such windows are likely necessary to separate various sections of the system in order to facilitate repair or replacement of parts of the lattice.

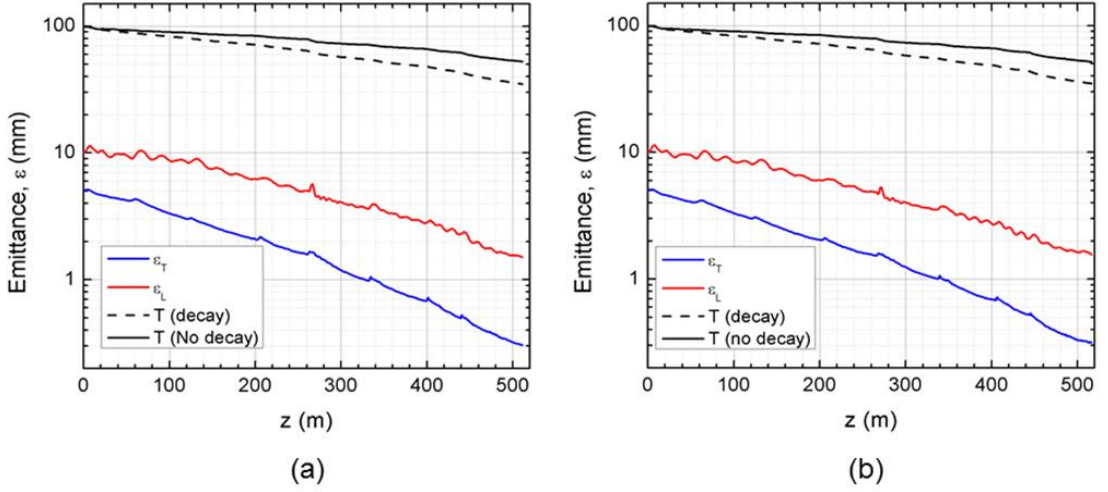

**Figure 6:** Simulation results of the performance of a tapered 6D cooling channel for a Muon Collider. The plot shows the evolution of the normalized rms emittances and transmissions vs. distance along the channel: (a) for a channel with vacuum rf cavities, and (b) a channel with gas filled cavities. The gas pressure is set at 34 atm at room temperature.

The cooling efficiency [9] of the lattice is often characterized in terms of the quality factor

$$Q_{6D} = \frac{d\epsilon_{6D}^N/ds}{dN/ds} \frac{N(s)}{\epsilon_{6D}^N}, \tag{4.1}$$

where $\epsilon_{6D}^N$ is the normalized six-dimensional emittance of the beam and N(s) is the number of surviving particles at arclength s. The Q factor compares the rate of change of emittance to the particle loss and under ideal conditions should remain constant through the lattice. In vacuum lattices Q starts off small due to losses from initial mismatching, then rises to a large value and finally falls as the emittance of the beam approaches its equilibrium value. Fig. 7 shows the evolution of Q along our hybrid 6D cooling channel. A glance at Fig. 7 indicates the importance of tapering as the cooling rate remains relatively flat with a maximum value Q≈7. The blue trace corresponds to an equivalent channel that uses vacuum rf cavities with LiH absorbers in every stage. It is worth noting that vacuum rf and hp rf channels achieve the same performance. This suggests that our hybrid approach is not degrading the cooling efficiency relative to the pure vacuum solution, thus making it a very promising choice.

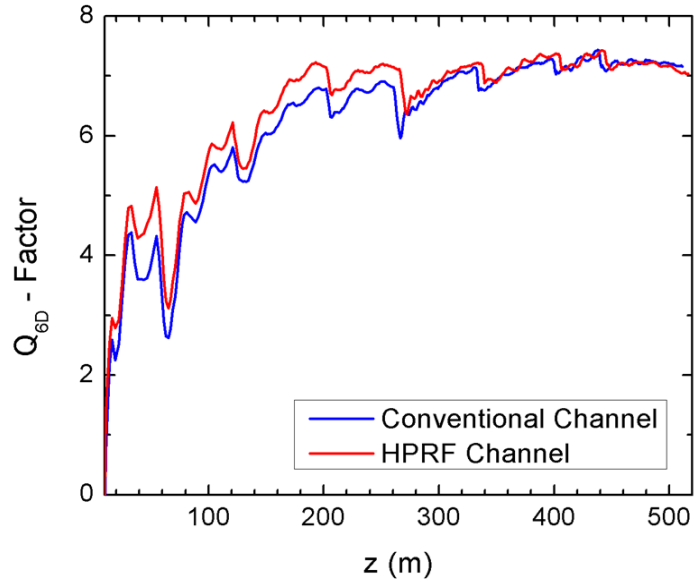

**Figure 7:** Quality factor, Q, versus z for a channel with vacuum and gas-filled rf cavities. Both channels achieve the same cooling efficiency along the channel.

Fig. 8 displays the final transverse emittance and overall transmission for different hydrogen gas pressures. As expected, an increase in the gas pressure degrades the ability of the channel to achieve micron scale emittances while simultaneously reducing the muon transmission. This is because the effect of multiple scattering in the absorber is now a bigger fraction of the overall angular beam divergence and thus a bigger contribution to the transverse emittance. Furthermore, the presence of very high pressure hydrogen everywhere in the lattice precludes placing the absorbers (as is possible in the vacuum lattice) at foci where $\beta_T$ is much lower than the average. While this effect on the beam emittance is minor at lower pressures <100 atm, at higher pressures, the cooling performance degrades substantially.

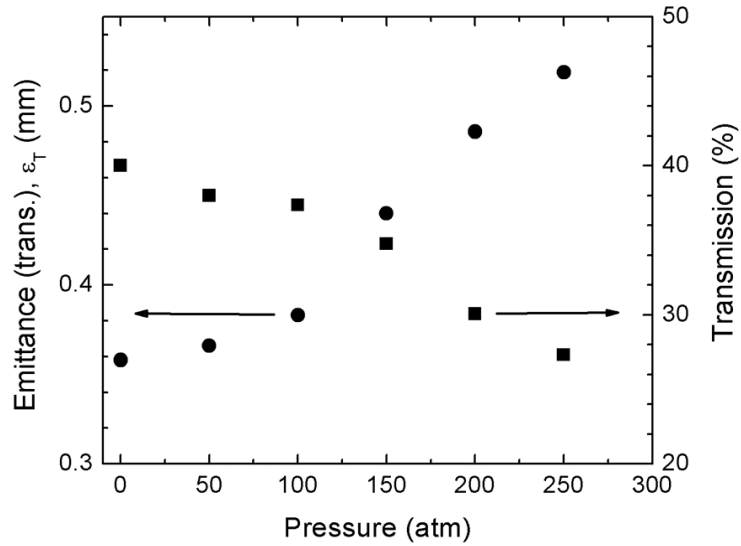

**Figure 8:** Performance vs. gas pressure. Dots show the emittance while the squares show the transmission. Hydrogen gas is assumed for all cases.

Fig. 9 compares the performance for two different gases, hydrogen and helium, both at 100 atm, with that of a vacuum channel. As expected from Fig. 8, a vacuum channel achieves the lowest possible emittance. On the other hand, hydrogen gas has the potential to cool the beam emittance to values that are at least 15% lower as compared to helium and thus will be the preferred choice between the two. Also, hydrogen suppresses breakdown at a voltage six times higher than helium [28]. Nitrogen is another possibility. It is not flammable, suppresses breakdown well, and its $L_R$ (dE/ds) may be large enough for initial cooling.

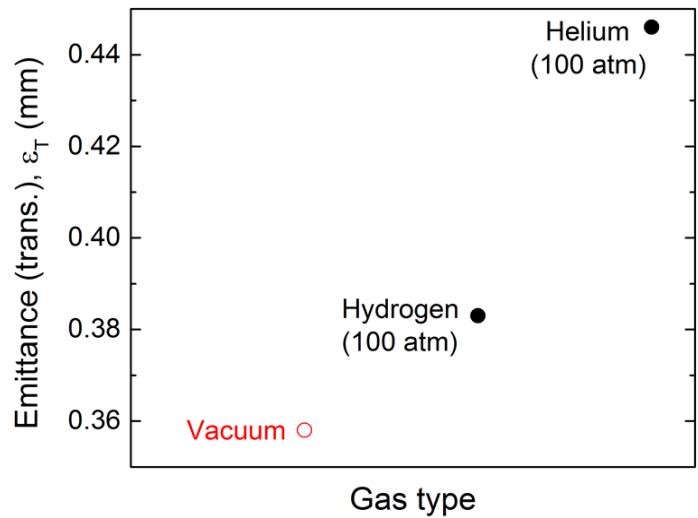

**Figure 9:** Performance vs. gas species. The pressure is assumed to be 100 atm for all cases.

Table 1: Main lattice parameters of a tapered rectilinear cooling channel. The first four stages have rf cavities at 325 MHz while the remaining stages have cavities at 650 MHz. Last two columns show the peak-axial-field and minimum beta function for each stage.

| Stage | Cell Length [m] | No. of Cells | Pipe Radius [cm] | rf Voltage [MV/m] | No. of rf Cav. | Coil Tilt [deg.] | Wedge Angle [deg.] | $B_{max}$ (T) | $\beta_{min}$ (cm) |
|---|---|---|---|---|---|---|---|---|---|
| 1 | 2.750 | 20 | 28.0 | 19.0 | 6 | 0.90 | 30 | 2.8 | 39.7 |
| 2 | 2.000 | 33 | 24.0 | 20.0 | 5 | 0.90 | 30 | 3.8 | 28.2 |
| 3 | 1.500 | 54 | 20.0 | 20.0 | 4 | 0.80 | 40 | 4.9 | 20.7 |
| 4 | 1.270 | 46 | 17.0 | 22.0 | 3 | 0.90 | 40 | 6.0 | 14.6 |
| 5 | 0.995 | 72 | 14.0 | 26.0 | 6 | 0.60 | 50 | 7.9 | 10.4 |
| 6 | 0.806 | 83 | 10.0 | 26.0 | 4 | 0.65 | 60 | 10.1 | 7.7 |
| 7 | 0.806 | 49 | 6.5 | 26.0 | 4 | 0.65 | 90 | 11.5 | 5.8 |
| 8 | 0.806 | 50 | 5.3 | 28.0 | 4 | 0.60 | 120 | 13.2 | 3.7 |
| 9 | 0.806 | 43 | 4.8 | 29.0 | 4 | 0.55 | 120 | 13.4 | 3.3 |

## 5. Summary

We have described a cooling technique that simultaneously reduces all six phase-space dimensions of a charged particle beam. We showed that such cooling can be achieved in a channel that consists of periodically inclined solenoids of alternating polarity, dense absorbers placed inside the solenoids, and rf cavities between them. We have discussed a possible scheme of using a hybrid approach wherein gaseous hydrogen provides protection of the rf cavities from breakdown in high magnetic fields and LiH absorbers are the primary energy loss medium. With the aid of numerical simulations, we showed that such a hp channel not only achieves the same transmission as vacuum designs, but also maintains its performance resulting in virtually the same 6D emittance reduction rate. Given the recent experimental studies that demonstrated that a hp gas-filled cavity can operate in a multi-tesla field without degradation, we believe that the hybrid concept presented here is a promising approach for ionization cooling channels for muon based applications.

### Acknowledgments

The authors are grateful to J. S. Berg, B. Freemire, R. B. Palmer, M. A. Palmer, H. Witte, and K. Yonehara for useful discussions. This work is supported by the U.S. Department of Energy, Contract no. DE-SC0012704.